\documentclass[aps,prl,reprint,superscriptaddress,nofootinbib,floatfix]{revtex4-2}

\usepackage{amsmath,amssymb,mathtools,bm}
\usepackage{graphicx}
\usepackage{xr-hyper}
\usepackage[hidelinks]{hyperref}

\newcommand{\Tr}{\operatorname{Tr}}
\newcommand{\Rea}{\operatorname{Re}}
\newcommand{\avg}[1]{\langle #1\rangle_\psi}
\newcommand{\topic}[1]{\par\smallskip\noindent\textit{#1.}\quad}
\graphicspath{{figures/}}
\begin{document}
\title{Reconstructing Pure-State OTOCs from Forward-Time Linear Response}
\author{Yu-Hang Jin}
\author{Xudong He}
\author{Yu Chen}
\thanks{ychen@gscaep.ac.cn}
\affiliation{Graduate School of China Academy of Engineering Physics, Beijing, 100193, China}
\date{\today}
\begin{abstract}
% Out-of-time-ordered correlators (OTOCs) diagnose quantum scrambling, yet their experimental reconstruction remains challenging. We show that the connected squared commutator of any pure state is exactly a normalized sum of squared commutator response slopes. These slopes resolve the component of the commutator orthogonal to the input state and are obtained using unitary string pulses, forward many-body evolution, and a fixed local readout, without inverse dynamics, ancillas, or simultaneous state copies. Restricting the pulses to a fixed Majorana weight produces an exact Krawtchouk filter and separates deterministic truncation bias from statistical sampling variance. Gaussian Fock states admit exact reconstruction formulas, including a single-weight calibration for two \(Z\) probes in the transverse-field Ising model. A complex-Haar benchmark predicts typical residuals of order \(\mathcal N_k^{-1/2}\), where \(\mathcal N_k=\binom Nk\), tested numerically in the Sachdev--Ye--Kitaev model and a nonintegrable Ising chain. The pulse ensemble thus becomes part of the reconstruction theory.
We show that the connected out-of-time-ordered correlator (OTOC) of any pure state can be reconstructed exactly from commutator linear responses alone. The protocol requires no time reversal, no ancilla, no double copy, and no weak measurement: a unitary string pulse, forward many-body evolution, and a single local readout suffice. An average over a complete basis of pulse generators yields the OTOC exactly. We further show that the measurement can be made sparse by restricting the pulses to a fixed operator weight, which turns the truncation into an analytically controlled spectral filter: the deterministic bias can be computed before the experiment and cleanly separated from the statistical sampling variance. For Gaussian states, the finite-weight response is exactly proportional to the full OTOC after a known combinatorial rescaling. For quantum chaotic systems, the typical residual is suppressed as a power law in the sector size. Exact-diagonalization benchmarks in the Sachdev-Ye-Kitaev model and a nonintegrable Ising chain confirm these predictions.
\end{abstract}
\maketitle

\topic{Introduction} --- Out-of-time-ordered correlators (OTOCs) quantify the growth of noncommutativity under many-body evolution \cite{larkin,roberts2015,kitaev,mss,xuswingle}. Their unusual operator ordering has motivated time-reversal protocols \cite{swingle,kastner2024} and ancilla-assisted interferometry \cite{zhu}. To bypass time reversal, several forward-time protocols have been developed, including randomized measurements \cite{vermersch}, classical shadows \cite{garcia,wang2024robustprocess}, double copy \cite{shen} and schemes based on quasiprobabilities \cite{halpern2017,kd}, sequential generalized measurements \cite{dressel}. These protocols have recently been demonstrated in  liquid- and solid-state nuclear magnetic resonance \cite{li,li2026}, trapped ions \cite{garttner,joshi,snapshot2024}, superconducting circuits \cite{mi,braumuller,google2025,wang2025}, and neutral atoms \cite{liang,xiang}.

Among these forward-time protocols, a particularly notable scheme is the randomized-measurement approach of Vermersch \textit{et al.} \cite{vermersch}, which reconstructs the OTOC by correlating measurement outcomes over randomly prepared initial states. Compared to time-reversal schemes, this approach is far less sensitive to control errors and decoherence, since it never requires backward evolution. Its only fundamental limitation is that state randomization restricts the direct output to the infinite-temperature trace OTOC.

Across these efforts, a common belief has emerged: that the OTOC cannot be directly accessed by commutator linear responses alone. Here we find otherwise. A conventional linear-response coefficient yields the expectation value of a commutator \cite{kubo}, whereas the connected squared commutator measures its fluctuations. This distinction appears to place pure-state OTOCs beyond linear response. However, for a pure state, these fluctuations are exactly the sum of squared off-diagonal matrix elements in the energy eigenbasis. The real question is therefore not whether linear response suffices in principle, but how the OTOC can be reconstructed from a family of commutator responses alone, while keeping both the prepared state and the final readout fixed.

We show that it can. Let \(A_t=[O_i(t),O_j]\) be the commutator of two local probes under forward evolution. For a pulse generator \(W_\mu\), the response \(R_\mu=\langle[W_\mu,A_t]\rangle_\psi\) is the expectation of the commutator between the pulse and the evolved probe. Equivalently, it is the overlap between the pulse direction \(W_\mu|\psi\rangle\) and the evolved vector \(A_t|\psi\rangle\). The complete set of these responses resolves the component of \(A_t|\psi\rangle\) that lies outside the initial ray. For any prescribed pure state, the connected squared commutator \(C(t)\) is exactly a normalized sum of squared responses over a complete basis of unitary pulses. The protocol uses unitary pulses, forward many-body evolution, and a single final readout per run; it requires no ancilla and no intermediate generalized measurement.

The reconstruction admits sparse pulse ensembles. Restricting the generators to a fixed operator weight keeps a polynomial number of settings and turns the truncation into an analytically controlled spectral filter. The deterministic bias can be computed before the experiment and cleanly separated from the statistical sampling variance. For Gaussian states, the finite-weight response is exactly proportional to the full OTOC after a known combinatorial rescaling. For chaotic systems, the typical residual is suppressed as a power law in the sector size. Exact-diagonalization benchmarks in the Sachdev-Ye-Kitaev model and a nonintegrable Ising chain confirm these predictions. This makes the choice of pulse ensemble part of the reconstruction theory.

\begin{figure}[t]
\centering
\includegraphics[width=\linewidth]{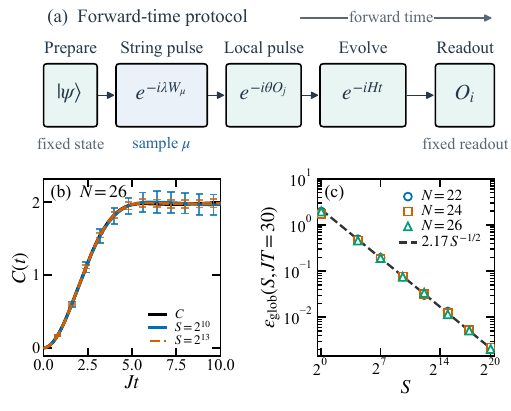}
\caption{\label{fig:global}Forward-time protocol and full-basis sampling.
(a) A fixed pure input receives a string pulse, a local pulse, forward evolution, and a fixed local readout. For each sampled \(\mu\), four pulse-sign settings determine \(R_\mu\) through Eq.~\eqref{eq:response}.
(b) Exact \(C(t)\) and ensemble-averaged estimators in middle-spectrum SYK\(_4\) at \(N=26\), with \(S=1024,8192\).
Bars show one standard deviation across 30 independent operator-sampling ensembles.
(c) Mean relative time-RMS error \(\epsilon_{\rm glob}\) against \(C\) at \(N=22,24,26\), evaluated per \(S\)-string estimator before averaging. The dashed line is the empirical guide \(2.17S^{-1/2}\), fitted jointly over \(1\le S\le2^{20}\). Errors use \(0.1\le Jt\le JT\), \(JT=30\).}
\end{figure}

\topic{Reconstruction and pulse sequence}
Let \(A_t=[O_i(t),O_j]\) for Hermitian probes labeled \(i,j\), with \(O_i(t)=e^{iHt}O_i e^{-iHt}\) under Hamiltonian \(H\) and \(\hbar=1\). The pure input \(\rho=|\psi\rangle\langle\psi|\) has expectation \(\avg{X}=\Tr(\rho X)\) for any operator \(X\). For a complete Hermitian operator basis \(W_\mu\) with \(\Tr(W_\mu W_\nu)=D\delta_{\mu\nu}\), where \(D\) is the Hilbert-space dimension and \(\delta_{\mu\nu}\) the Kronecker delta, define \(R_\mu=\avg{[W_\mu,A_t]}\). Then
\begin{equation}
 C(t)=-\avg{A_t^2}+\avg{A_t}^{\,2}
 =\frac{1}{2D}\sum_\mu R_\mu^2 .
 \label{eq:global}
\end{equation}
Indeed, \(R_\mu=\Tr(W_\mu[A_t,\rho])\), and \([A_t,\rho]\) is Hermitian. Parseval's identity gives \(\sum_\mu R_\mu^2=D\Tr([A_t,\rho]^2)\). Using \(\rho^2=\rho\) gives Eq.~\eqref{eq:global}. Equivalently, \(C=\|(I-\rho)A_t|\psi\rangle\|^2\). In an orthonormal energy eigenbasis \(\{|m\rangle\}\), the input \(|n\rangle\) gives \(\sum_{m\ne n}|\langle m|A_t|n\rangle|^2\). For involutory probes \(O_i^2=O_j^2=I\), with \(I\) the identity, let \(F=\avg{O_i(t)O_jO_i(t)O_j}\) be the four-point OTOC. Then \(2(1-\Rea F)=C+|\avg{A_t}|^2\). The mean commutator follows from the same local pulse with no string pulse.

The laboratory sequence is state preparation, \(U_\mu=e^{-i\lambda W_\mu}\), \(V_j=e^{-i\theta O_j}\), evolution under \(H\) for time \(t\), and a readout of \(O_i\) [Fig.~\ref{fig:global}(a)]. The pulse angles are \(\lambda,\theta\). The measured signal is
\begin{equation}
 s_\mu(t;\lambda,\theta)=\avg{U_\mu^\dagger V_j^\dagger O_i(t)V_j U_\mu}.
 \label{eq:signal}
\end{equation}
It satisfies
\begin{equation}
 \begin{aligned}
 R_\mu(t)&=\left.\partial_\lambda\partial_\theta s_\mu(t;\lambda,\theta)\right|_{0,0}\\
 &=\frac{\sum_{\sigma,\tau=\pm1}\sigma\tau\,s_\mu(t;\sigma\lambda,\tau\theta)}
 {\sin(2\lambda)\sin(2\theta)} .
 \end{aligned}
 \label{eq:response}
\end{equation}
The derivative identity holds generally. The finite-angle equality holds for \(W_\mu^2=O_j^2=I\) and nonzero denominators. The signs of the pulses select the mixed response, while the many-body evolution always runs forward. The protocol uses one state preparation at a time and requires neither an ancilla nor intermediate generalized readout. With independent uniform draws from the \(D^2\) strings, the sample mean of \(DR_\mu^2/2\) is unbiased for \(C\) when the slopes are known. With finite readout shots, two independent estimates of the same slope can be multiplied to keep its square unbiased.

\begin{figure}[t]
\centering
\includegraphics[width=\linewidth]{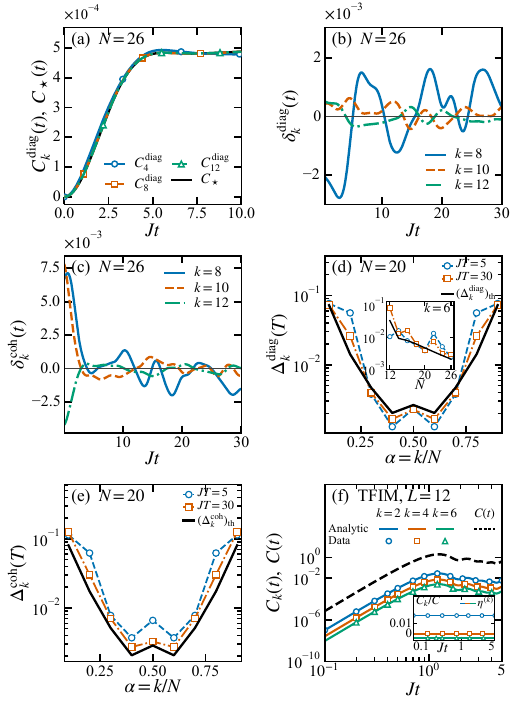}
\caption{\label{fig:sectors}\label{fig:twirling}Complete fixed-weight sums in SYK\(_4\) and the TFIM.
(a) \(C_k^{\rm diag}(t)\) for \(k=4,8,12\) and the sector reference \(C_\star(t)\) at \(N=26\).
(b),(c) Signed diagonal and coherent residuals for \(k=8,10,12\) at \(N=26\).
(d),(e) Residual time RMS versus \(\alpha=k/N\) at \(N=20\); inset of (d): size dependence at \(k=6\). Circles and squares use \(JT=5,30\), respectively, with \(0.1\le Jt\le JT\). Black curves give the complex-Haar estimates (subscript \({\rm th}\)).
(f) Open TFIM at \(L=12\), \(J=h=1\), in a near-center Fock eigenstate, with \(Z\) probes at sites 6 and 8. Symbols are complete-sector sums, colored curves are \(\eta_2^{(k)}C(t)\) for \(k=2,4,6\), and the dashed black curve is \(C(t)\). The inset compares \(C_k/C\) with \(\eta_2^{(k)}\equiv\eta^{(k)}\).}
\end{figure}

\topic{Fixed Majorana weight}
For \(L\) fermionic modes there are \(N=2L\) Majoranas \(\gamma_a\), \(D=2^L\), and \(\{\gamma_a,\gamma_b\}=2\delta_{ab}I\). Let \([N]=\{1,\ldots,N\}\) and \(\mu=\{\mu_1<\cdots<\mu_k\}\subseteq[N]\). The strings \(W_\mu=i^{k(k-1)/2}\gamma_{\mu_1}\cdots\gamma_{\mu_k}\) are Hermitian involutions. With \(\mathcal S_k=\{\mu:|\mu|=k\}\) and \(\mathcal N_k=\binom Nk\), define the channel \(\Phi_k\) and the sector response \(C_k\) by
\begin{equation}
 \Phi_k(X)=\frac{1}{\mathcal N_k}\sum_{\mu\in\mathcal S_k}W_\mu X W_\mu,
 \quad C_k=\frac{1}{2\mathcal N_k}\sum_{\mu\in\mathcal S_k}R_\mu^2 .
 \label{eq:fixed}
\end{equation}
Exchanging strings of weights \(k\) and \(q=|\nu|\) with \(s=|\mu\cap\nu|\) shared indices gives the sign \((-1)^{kq-s}\). Thus \(\Phi_k(W_\nu)=\kappa_{kq}W_\nu\), with
\begin{equation}
 \kappa_{kq}=\frac{1}{\mathcal N_k}\sum_s
 (-1)^{kq-s}\binom qs\binom{N-q}{k-s}.
 \label{eq:kappa}
\end{equation}
This is the normalized Krawtchouk eigenvalue, including the Clifford exchange sign. Binomial coefficients vanish for lower entries outside their integer range.

For an input with fermion-parity eigenvalue \(p_\psi=\pm1\), \(\rho_k=\Phi_k(\rho)\) lies in the sector of parity \((-1)^kp_\psi\) and dimension \(d=D/2\). Let \(\Pi_k=\mathbb I_{(-1)^kp_\psi}/d\), where \(\mathbb I_{(-1)^kp_\psi}\) is the identity on that sector, so \(\Tr\Pi_k=1\). The exact Hilbert--Schmidt distance is
\begin{equation}
 \|\rho_k-\Pi_k\|_2^2=\frac{1}{D}
 \sum_{\nu\notin\{\varnothing,[N]\}}\kappa_{k,|\nu|}^2|\avg{W_\nu}|^2 .
 \label{eq:distance}
\end{equation}
If \(A_t\) has definite parity, only even \(k\) for even \(A_t\) or odd \(k\) for odd \(A_t\) contributes. Set \(z_\mu=\avg{A_tW_\mu}\) and \(|v\rangle=A_t|\psi\rangle\). The expansion \(R_\mu^2/2=|z_\mu|^2+\Rea z_\mu^2\) yields
\begin{equation}
 \begin{aligned}
 C_k&=C_k^{\rm diag}+C_k^{\rm coh},\\
 C_k^{\rm diag}&=\langle v|\rho_k|v\rangle,\quad
 C_k^{\rm coh}=\frac{1}{\mathcal N_k}\Rea\sum_{\mu\in\mathcal S_k}z_\mu^2 .
 \end{aligned}
 \label{eq:split}
\end{equation}
The diagonal term measures the projection of the twirled state onto the probe direction. The coherent term retains the phases of the same amplitudes. The parity-sector reference is \(C_\star=\langle v|v\rangle/d\), so the connected target is \(dC_\star-|\avg{A_t}|^2\). Estimation via the parity-sector reference therefore requires this subtraction. The exact Gaussian reconstruction below targets \(C\) directly.

\topic{Truncation and sampling errors}
Draw \(S\) strings independently and uniformly from \(\mathcal S_k\) with replacement. Write \(Y_\mu=R_\mu^2/2\), let \(\widehat C_{k,S}\) be its sample mean, and use \(\|f\|_T^2=\int_{0^+}^T |f(t)|^2\,dt\) with the isolated zero-time point omitted. For any fixed comparison curve \(B(t)\),
\begin{equation}
 \begin{aligned}
 \mathbb E\|\widehat C_{k,S}-B\|_T^2
 ={}&\|C_k-B\|_T^2\\
 &+\frac{1}{S}\int_{0^+}^{T}
 \operatorname{Var}_{\mu\in\mathcal S_k}Y_\mu(t)\,dt .
 \end{aligned}
 \label{eq:variance}
\end{equation}
Here \(\mathbb E\) denotes the average over operator-string sampling at fixed state and Hamiltonian. For \(B=C_\star\), the deterministic floor includes the coherent term and its interference with the diagonal residual. Its value depends on the state and probe. The identity also holds on a fixed time grid with quadrature weights. It accounts for string sampling with precise slopes. Readout shots and pulse errors enter separately.

We define \(\delta_k^{\rm diag}=(C_k^{\rm diag}-C_\star)/C_\star\) and \(\delta_k^{\rm coh}=C_k^{\rm coh}/C_\star\) at nonzero signal, and define their time RMS by
\begin{equation}
 \Delta_k^x(T)=\left[\frac{1}{T}\int_{0^+}^{T}(\delta_k^x(t))^2\,dt\right]^{1/2},
 \quad x\in\{\mathrm{diag},\mathrm{coh}\}.
 \label{eq:delta-rms}
\end{equation} Complex-Haar statistics give a reference for typical fixed-weight residuals. Averaging Eq.~\eqref{eq:distance} over pure states in the source parity sector gives \(\big[d^2(1+\delta_{2k,N})/\mathcal N_k-1\big]/[d(d+1)]\). A second moment model for the relative orientation of the probe direction yields a diagonal relative RMS of order \(\mathcal N_k^{-1/2}\) for fixed \(k\) and large \(N\). A phase-sensitive Gram-matrix moment gives the coherent benchmark with the same fixed-\(k\) order. The factor at half weight follows from complementary strings, which act proportionally within a parity sector. These models predict a scale for the full-sector residual of a typical state. The residual itself is deterministic for a fixed state and time. The Supplemental Material (SM) \cite{supplement} derives both moments and gives exact expressions for testing the isotropy assumptions.

\begin{figure}[t]
\centering
\includegraphics[width=\linewidth]{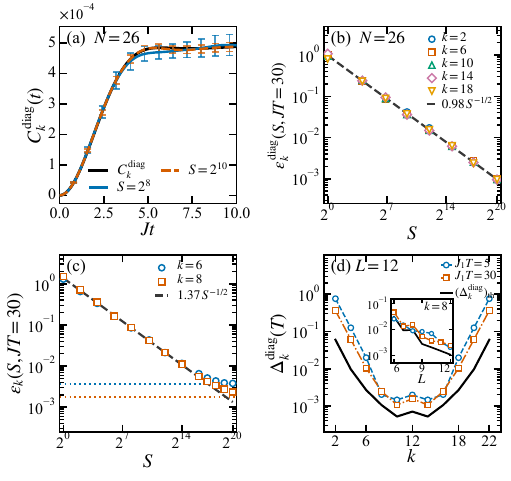}
\caption{\label{fig:sampling}\label{fig:fixedsampling}Sampling and truncation. Panels (a)--(c) use middle-spectrum SYK\(_4\) at \(N=26\).
(a) The \(k=12\) diagonal response at \(S=256,1024\) and its complete-sector value. Bars as in Fig.~\ref{fig:global}(b).
(b) Relative diagonal errors \(\epsilon_k^{\rm diag}\) against \(C_k^{\rm diag}\) for \(k=2,6,10,14,18\); the common \(0.98S^{-1/2}\) guide is fitted jointly over \(1\le S\le2^{20}\).
(c) Complete-response errors \(\epsilon_k\) at \(k=6,8\) relative to \(C_\star\). Dotted lines are exact sector floors. The common \(1.37S^{-1/2}\) guide retains the joint calibration from \(k=6,8,12\) over \(1\le S<2^{12}\). Errors in (b),(c) are computed per estimator before averaging over 30 ensembles, with \(0.1\le Jt\le JT\), \(JT=30\).
(d) Diagonal residual RMS versus Majorana weight in the nonintegrable Ising chain, Eq.~\eqref{eq:nnn-ising}, at \(L=12\). The inset varies \(L=6,\ldots,12\) at \(k=8\). Circles and squares use \(J_1T=5,30\); black solid curves in the main panel and inset give the complex-Haar prediction \((\Delta_k^{\rm diag})_{\rm th}\), SM Eq.~\eqref{SM-eq:supp-nnn-haar-rms}. The \(Z\)-probe sites follow the same rule as in SYK, now with \(L\) spins.}
\end{figure}

\topic{Exact Gaussian calibration}
Use an adapted Fock basis in which the Gaussian input is \(|\psi\rangle=|\mathbf n\rangle\). Here \(\mathbf n=(n_1,\ldots,n_L)\) and \(\mathbf m=(m_1,\ldots,m_L)\) are initial and final occupation vectors, with \(n_a,m_a\in\{0,1\}\) for mode \(a=1,\ldots,L\). Their occupation distance is \(d_{\rm occ}(\mathbf m,\mathbf n)=\sum_{a=1}^L|m_a-n_a|\). For \(r=0,\ldots,L\), the total squared transition amplitude to states at distance \(r\) is
\begin{equation}
 u_r(t)=\sum_{\mathbf m:\,d_{\rm occ}(\mathbf m,\mathbf n)=r}
            |\langle\mathbf m|A_t|\mathbf n\rangle|^2 .
 \label{eq:occupation-shell}
\end{equation}
In this basis a Majorana string maps the input to one occupation state up to phase. The fraction of weight-\(k\) strings reaching a specified final state at distance \(r\) is
\begin{equation}
 \eta_r^{(k)}=\frac{2^r}{\binom{2L}{k}}
 \binom{L-r}{(k-r)/2}.
 \label{eq:count}
\end{equation}
The coefficient is zero unless \((k-r)/2\) is an integer between \(0\) and \(L-r\). Each changed mode requires one Majorana from its pair, while \((k-r)/2\) other modes contribute both. Real orthogonal rotations of the Majoranas mix strings only within the same weight and leave the sum of squared slopes invariant. Thus the counting applies to the experimental basis. For each \(r>0\), swapping the Majorana choice on one changed mode reverses the sign of the squared complex amplitude in \(C_k^{\rm coh}\). For \(r=0\) the string eigenvalues are real and the coherent term cancels the diagonal contribution. Consequently
\begin{equation}
 C_k(t)=\sum_{r=1}^L \eta_r^{(k)}u_r(t),
 \quad C(t)=\sum_{r=1}^L u_r(t).
 \label{eq:shell}
\end{equation}
The sums contain \(L\) occupation-distance terms; TFIM \(ZZ\) probes involve only \(r=0,2\), with \(r=0\) removed by the connected subtraction.
Weights \(k=r\) for the allowed positive \(r\) give a triangular system with a nonzero diagonal. Its inverse supplies a known linear combination of complete fixed-weight responses. The inversion coefficients also determine how measurement noise is amplified.

For the open TFIM, \(H=-J\sum_{j=1}^{L-1}X_jX_{j+1}-h\sum_{j=1}^L Z_j\), where \(X_j,Z_j\) are site Pauli operators and \(J,h\) are the exchange and field strengths. The Jordan--Wigner transformation makes \(H\) quadratic \cite{pfeuty1970ising}. The commutator \([Z_i(t),Z_j]\) is bilinear in the Majoranas and can change only zero or two occupations. Equation~\eqref{eq:shell} then gives \(C_k=\eta_2^{(k)}C\), with \(\eta_2^{(k)}\equiv\eta^{(k)}\), exactly for even \(2\le k\le 2L-2\). The relation holds for any pure Gaussian input, including quasiparticle Fock eigenstates. The supplemental \(XX\) and \(XZ\) examples illustrate when several weights are needed.

\topic{Numerical benchmarks}
We diagonalize SYK\(_4\) \cite{sachdev,kitaev,maldacena},
\(H=\tfrac14\sum_{a<b<c<e}J_{abce}\gamma_a\gamma_b\gamma_c\gamma_e\),
with independent centered Gaussian couplings of variance \(6J^2/N^3\), where \(J\) sets the SYK energy scale.
We use middle-spectrum eigenstates, \(N\le26\), and \(Z\) probes at \(i=\lfloor(L-2)/2\rfloor+1\) and \(j=i+2\), with \(L=N/2\) and one-based site labels.
Figure~\ref{fig:global}(b,c) shows convergence to the direct OTOC and the \(S^{-1/2}\) scaling of its relative sampling error. In Figs.~\ref{fig:global}(c) and \ref{fig:sampling}(b,c), \(\epsilon_{\rm glob},\epsilon_k^{\rm diag},\epsilon_k\) denote \(\|\widehat Q-B\|_T/\|B\|_T\) for the sampled curves \(\widehat Q\) with references \(B=C,C_k^{\rm diag},C_\star\), respectively. Errors are evaluated separately for each \(S\)-string estimator before averaging over 30 independent operator-sampling ensembles, using \(0.1\le Jt\le JT\) with \(JT=30\). All plotted times use \(\hbar=1\); \(0^+\) in the integrals denotes omission of the zero-time point, with the numerical grid starting at \(Jt=0.1\).

To test fixed-weight truncation separately, we sum all strings in each sector.
The diagonal signal tracks \(C_\star\), its signed residual fluctuates about zero, and the coherent residual changes sign over time [Figs.~\ref{fig:sectors}(a)--(c)]. The RMS residuals fall toward half weight, with a local enhancement at \(k=N/2\) captured by the complementary-string term in the Haar reference [Figs.~\ref{fig:sectors}(d),(e)]. Figure~\ref{fig:sectors}(f) tests the distinct Gaussian prediction \(C_k=\eta_2^{(k)}C\) on a near-center Fock eigenstate.

Figure~\ref{fig:sampling} restores finite sampling. The diagonal response converges to its sector value, with errors following the sampling law in Eq.~\eqref{eq:variance}. For the complete response the relative floor is \(\|C_k-C_\star\|_T/\|C_\star\|_T\). The \(k=6\) error approaches its floor, while \(k=8\) has a lower floor and stays above it over the sampled range. The nonintegrable benchmark is the open chain
\begin{equation}
 H=-J_1\sum_{j=1}^{L-1}X_jX_{j+1}
   -J_2\sum_{j=1}^{L-2}X_jX_{j+2}
   -h\sum_{j=1}^{L}Z_j,
 \label{eq:nnn-ising}
\end{equation}
with \(J_2/J_1=0.5\), \(h/J_1=1.05\), Majorana fixed-weight pulses, and a near-center eigenstate in the even-parity sector. These data show a comparable suppression of diagonal residuals with weight and system size [Fig.~\ref{fig:sampling}(d)]. The Supplemental Material provides size and energy sweeps and stabilizer-state benchmarks.

\topic{Implementation and conclusion}
Fixed Majorana weight controls the population of pulse settings. The Jordan--Wigner image of a fixed-weight monomial can still have support proportional to system size. A direct Pauli-string rotation with Pauli support \(\ell\) (the number of nonidentity factors) uses \(2(\ell-1)\) entangling gates with suitable connectivity. Odd-weight strings can be implemented as Pauli rotations in the qubit encoding. Physical fermionic implementations need control compatible with parity superselection. The string-pulse budget per time point for \(S\) sampled settings is then \(O(SN)\), prior to multiplying by pulse settings, repetitions, and routing. A bounded relative sampling prefactor gives \(S=O(\epsilon_{\rm stat}^{-2})\) for a prescribed statistical error \(\epsilon_{\rm stat}\). State preparation, readout precision, and calibration determine the experimental cost of attaining that error.

The squared response identity reconstructs the connected OTOC of any prescribed pure state from forward-time measurements. Majorana-weight restriction preserves enough structure to quantify what is lost: in Gaussian Fock states the missing contributions are recovered by explicit combinatorial coefficients, and in chaotic benchmarks the residuals can be compared with a finite-size statistical prediction. This makes the choice of pulse ensemble part of the reconstruction theory.

\par\smallskip
\begin{samepage}
\begin{acknowledgments}
This work is supported by the National Key R\&D Program of China (Grant No.\ 2022YFA1405302), the National Natural Science Foundation of China (Grant No.\ 12675031 and No.\ 12174358), and NSAF (Grant No.\ U2330401).
\end{acknowledgments}
\end{samepage}

\end{document}